# Probing Hidden Symmetry and Altermagnetism with Sub-Picometer Sensitivity via Nonlinear Transport

Subin Mali[1+], Yufei Zhao[3+], Yu Wang[1], Saugata Sarker[2], Yangyang Chen[1], Zixuan Li[1], Jun Zhu[1], Ying Liu[1], Venkatraman Gopalan[1,2], Binghai Yan [1,3*] and Zhiqiang Mao [1,2*]

[1] Department of Physics, Pennsylvania State University

[2] Department of Material Science and Engineering, Pennsylvania State University

[3] Department of Condensed Matter Physics, Weizmann Institute of Science

## Abstract

X-ray and neutron diffraction are foundational tools for determining crystal structures, but their resolution limits can lead to misassignments—especially in materials with subtle distortions or competing phases. Here, we demonstrate the use of nonlinear transport as a complementary approach to uncover hidden crystal symmetries, using the strongly correlated $Ca_3Ru_2O_7$ as a case study. Below 48 K ($T_S$), where the magnetic moments of the antiferromagnetic phase reorient from the $a$- to the $b$-axis, leading to a pseudogap opening, our measurements, with support of DFT, reveal a previously overlooked lower-symmetry phase. This is manifested by the emergence of longitudinal nonlinear resistance (NLR) along the $b$-axis below $T_S$, providing direct evidence of combined translational and time-reversal symmetry breaking. This response also suggests a transformation from a conventional antiferromagnet into an altermagnet. The lower-symmetry phase arises from a subtle lattice distortion (~0.1 pm) associated with the magnetic transition at $T_S$, below the detection limit of conventional diffraction. Moreover, this NLR below $T_S$ is accompanied by a nonlinear Hall effect, both of which are enhanced by the large quantum metric associated with Weyl chains near the Fermi surface. Our findings demonstrate nonlinear transport as a sensitive probe of hidden symmetry breaking and altermagnetism, complementing and extending beyond the reach of traditional diffraction and spectroscopic techniques.

[+]S.M. & Y.Z. made equal contributions to this work.

[*]emails: binghai.yan@psu.edu; zim1@psu.edu

**Introduction**

Symmetry and symmetry breaking lie at the heart of modern condensed matter physics. In particular, Ginzburg-Landau theory frames phase transitions in terms of spontaneous symmetry breaking and the emergence of order parameters[1]. Accurate determination of a material's crystal symmetry is therefore essential for understanding its physical properties and identifying its ground state. Conventionally, X-ray and neutron diffraction are foundational tools for crystallographic analysis. However, these methods have resolution limits which can obscure subtle structural distortions and lead to incorrect structural assignments, particularly in complex materials with competing phases or correlated electronic states. A notable example is the massive Dirac material $BaMnSb_2$, which was originally assigned a tetragonal space group based on neutron diffraction data[2,3]. However, recent studies employing scanning transmission electron microscopy (STEM), second harmonic generation (SHG), and density functional theory (DFT) have shown that $BaMnSb_2$ actually adopts an orthorhombic polar structure with a subtle distortion (~0.3 Å)[4], thereby uncovering a bulk spin-valley-locked Dirac state[4] and a nonlinear Hall effect (NLHE)[5]. These approaches compellingly demonstrate that resolving hidden symmetry breaking can uncover new topological phases; however, they often require large-scale facilities, specialized instrumentation, and coordinated experimental–theoretical analysis. Given that identifying hidden symmetry breaking likely leads to the discovery of new topological phases, there is a clear need for alternative, more accessible approaches to detect subtle symmetry changes and associated topological phenomena.

In this work, we demonstrate a simple yet powerful alternative: a metrological method for detecting hidden symmetry breaking through nonlinear transport measurements. Recent advances have established nonlinear transport measurements as a sensitive probe of band topology and quantum geometry, particularly in non-centrosymmetric material [6–13]. We apply this method to $Ca_3Ru_2O_7$ (CRO), a prototypical strongly correlated oxide known for its rich phase diagram and emergent phenomena[14–16], including colossal magnetoresistance[15,17], spin-valve effects[14], metamagnetic textures[18,19], polar semimetal behavior[20,21], and massive Dirac fermions[22]. Upon cooling, the system undergoes two sequential magnetic transitions: a paramagnetic (PM) to A-type antiferromagnetic (AFM-a) phase at $T_N$ = 56 K, followed by a spin reorientation into the AFM-b phase at $T_s$ = 48 K (see Fig. 1g). The magnetic transition at $T_s$ is accompanied by the opening of a pseudogap and a sharp increase in resistivity, signaling strong coupling between spin, lattice, and charge degrees of freedom. Recent Raman spectroscopy further reveals a switch in dominant infrared-active phonon modes across this transition, modulating both pseudogap formation and hopping pathways[23].

Despite decades of neutron diffraction refinements assigning the CRO structure to the *Bb2₁m* space group at all temperatures, our nonlinear transport measurements, supported by first-principles calculations[24–26], uncover a lower-symmetry *Pn2₁a* structure that emerges below $T_s$. This structural transition is driven by an extremely small lattice distortion (~0.001 Å), induced by the spin reorientation—below the resolution limit of conventional neutron scattering. Importantly, this symmetry lowering is not just a structural curiosity: it reclassifies the AFM-b phase as a unique altermagnet[27–29], which breaks the combined translational ($\tau$) and time-reversal ($\mathcal{T}$) symmetry, $\tau\mathcal{T}$. The nonrelativistic spin split in the CRO's altermagnetic phase is negligibly small due to the smallness of distortion and the anomalous Hall effect vanishes because of extra symmetry constraint, hindering the identification of the altermagnet state by the conventional spectroscopic and linear magneto transport measurements.

However, in our nonlinear transport measurements on CRO, we observed a pronounced longitudinal nonlinear resistance (NLR) along the polar *b*-axis in the AFM-b phase. This response serves as direct evidence of $\tau\mathcal{T}$-breaking, consistent with altermagnetic symmetry class. Furthermore, we found this longitudinal nonlinear response is accompanied by a NLHE. Both longitudinal and transverse nonlinear effects are significantly enhanced by the large quantum metric[30–34] associated with Weyl chains[25] near the Fermi surface, revealing a complex interplay between crystal symmetry, electronic topology, and transport in CRO (see Fig. 1). Taken together, our results establish nonlinear transport as a highly sensitive probe of hidden symmetry breaking in non-centrosymmetric materials. This approach opens an accessible path to identifying elusive topological and altermagnetic phases in a broad class of emerging quantum materials.

**Results**

**Crystal symmetry, electronic topology and nonlinear transport**

As noted above, prior neutron diffraction structure refinements show that CRO adopts the polar space group *Bb2₁m* [35] where RuO₆ octahedra form bilayer structures with ferromagnetic coupling inside the bilayer and AFM coupling in between bilayers. Below $T_S$, the corresponding magnetic space group of AFM-b is $P_Cna2_1$ (No. 33.154), characterized by a two-fold screw-rotation $\{C_{2y}|\left(0,\frac{b}{2},0\right)\}$, two glide reflections $G_x = \{M_x|\left(\frac{a}{2},\frac{b}{2},\frac{c}{2}\right)\}$ and $G_z = \{M_z|\tau\}$, and $\tau\mathcal{T}$ with $\tau = \left(\frac{a}{2},0,\frac{c}{2}\right)$. Here $\tau\mathcal{T}$ connects two spin sublattices and renders the *Bb2₁m* phase an ordinary AFM. Recent DFT calculations[24,26] suggests a lower-symmetry crystal structure with the space group *Pn2₁a* as the ground state of AFM-b. As illustrated in Fig. 1b, the RuO₆ octahedra breath in and out weakly by ~ 0.001 Å in the *Pn2₁a* structure. While this distortion is subtle and undetectable by neutron scattering, it breaks the $\tau\mathcal{T}$ symmetry, thereby converting an AFM to an altermagnet according to symmetry principles. This symmetry-reduction alters the band structure and subsequently induces nonlinear transport effects, as discussed in the following.

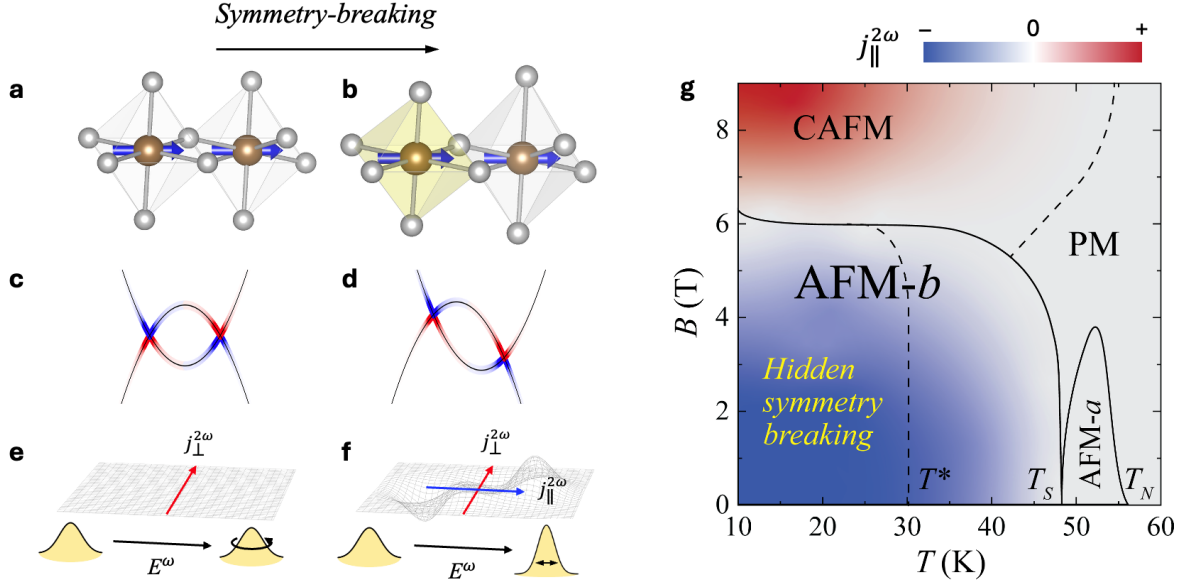

**Figure 1. Schematics of symmetry-breaking from *Bb2₁m* to *Pn2₁a* phases.** (a-b) The structure distortion leads to Weyl band tilting (c-d). Weyl bands generate giant quantum metric whose sign is represented by blue ("−") and red ("+") and thus lead to net quantum metric at the Fermi energy in the tilted case of (d). (e) The *Bb2₁m* exhibits only nonlinear Hall effect ($j_\perp^{2\omega}$) from the Berry curvature, illustrated by the self-rotation of wave packet in the presence of an electric field ($E^\omega$), because of $\tau T$-symmetry. (f) The *Pn2₁a* exhibits both nonlinear Hall effect and nonlinear longitudinal resistance ($j_\parallel^{2\omega}$), mainly caused by the quantum metric illustrated by the wave packet deformation, because of $\tau T$-breaking induced Weyl band tilting. (g) The schematic magnetic phase diagram of $Ca_3Ru_2O_7$ for the magnetic field applied along the *b*-axis, adapted from ref. [36]. The blue and red background illustrate the longitudinal nonlinear current measured in our experiment (see below).

Our DFT calculations[25] show the existence of Weyl chains near the Fermi energy for both the *Bb2₁m* and *Pn2₁a* phases. As illustrated in Fig. 1c-d, Weyl bands are symmetric around the M point (the Brillouin zone boundary) for the *Bb2₁m* phase because of the $\tau T$ symmetry while they turn tilted for the *Pn2₁a* phase because of $\tau T$-breaking. We note that Weyl points extend continuously in glide planes ($G_x$ and $G_z$) and form inter-connected Weyl chains. Weyl bands generate giant quantum metric near Weyl points as shown in Fig. 1c-d where blue and red represent opposite signs of quantum metric distribution. For symmetric Weyl bands in *Bb2₁m*, the net quantum metric is strictly zero because of $\tau T$-induced cancellation. For the asymmetric Weyl bands of *Pn2₁a*, the net quantum metric is significant despite weak tilting, leading to a large NLR along the polar axis *b*. Additionally, the calculated Fermi pockets in both phases agree with carrier densities determined by our Hall measurement ($n_e = 7.0 \times 10^{18}$ cm$^{-3}$ and $n_h = 7.3 \times 10^{18}$ cm$^{-3}$) (see supplementary Note 1 and supplementary Fig. S1).

The $Bb2_1m$ and $Pn2_1a$ phases exhibit qualitatively different symmetry in nonlinear transport although both are inversion breaking and present nonlinear transport. Because of $\tau\mathcal{T}$ symmetry which is equivalent to $\mathcal{T}$ to constrain the nonlinear response, $Bb2_1m$ exhibits only a NLHE[37] induced by Berry curvature dipole (BCD) (Fig. 1e). Free from the $\tau\mathcal{T}$-constraint, $Pn2_1a$ exhibits both NLHE due to BCD and quantum metric dipole (QMD) and longitudinal NLR caused by QMD and a Drude term[34], as illustrated in Fig. 1f. We summarize the symmetry-constrained nonlinear conductivities in Table I for both $Bb2_1m$ and $Pn2_1a$. According to our calculations [25], QMD is several orders of magnitude larger than BCD and Drude contributions here. Therefore, we expect that NLHE and NLR will increase quickly after entering the AFM-b phase and the existence of NLR is the hallmark to identify the $Pn2_1a$ phase and the $\tau\mathcal{T}$-breaking altermagnet. However, conventional methods cannot distinguish between these two phases. For example, optical SHG shows the same nonlinear susceptibilities for both phases (see supplementary Table 1). Likewise, the nonrelativistic spin splitting, which is smaller than 1 meV, and anomalous Hall effect, which vanishes due to the coexistence of glide mirror symmetries $G_x$ and $G_z$ in CRO, fail to discriminate the altermagnet from an ordinary antiferromagnet. Importantly, the combined $\tau\mathcal{T}$-symmetry forbids both the altermagnetic phase and the presence of longitudinal NLR. Thus, detection of NLR provides a clear signature of $\tau\mathcal{T}$-breaking, compatible with the noncentronsymmetric altermagnetic state in CRO, direct observation of spin-split bands will be required for definitive confirmation.

**Table 1** Symmetry constrained second-order conductivity tensors in $Bb2_1m$ and $Pn2_1a$ phase of $Ca_3Ru_2O_7$ in the AFM-b state. Note that the longitudinal term $\sigma_{b;bb}$ is forbidden in $Bb2_1m$ but allowed in $Pn2_1a$. In $Bb2_1m$, the nonlinear Hall effect is contributed by the Berry curvature dipole (BCD). In $Pn2_1a$, the nonlinear transport can be contributed by the nonlinear Drude, BCD and quantum metric dipole (QMD).

| Space group | $Bb2_1m$ | $Pn2_1a$ |
|---|---|---|
| Second-order tensors [AFM-b ($T < T_S$)] | $\sigma_{b;aa}, \sigma_{b;cc}$ (BCD) | $\sigma_{b;aa}, \sigma_{b;cc}$ (Drude, BCD, QMD) $\sigma_{b;bb}$ (Drude, QMD) |

**Nonlinear longitudinal transport measurements**

To find if the AFM-b phase exhibits NLR along the *b*-axis as the theory predicted, we fabricated a microscale Hall bar device with its long axis aligned along the *b*-axis (denoted as S1) using a combination of photolithography and focused ion beam (FIB) deposition (see Methods) and performed nonlinear transport measurements on the device with a current applied along the *b*-axis. Microscale devices offer two key advantages: They enhance any existing nonlinear response by enabling the application of larger electric fields without significant heating, while also minimizing the influence of multiple domains. Given the typical domain size in CRO is approximately 20 μm, our devices were fabricated on single-domain regions. The crystallographic axes of the CRO crystals used in the devices were identified via SHG measurements (Supplementary Note 2 and Supplementary Fig. S2) and verified through magnetoresistance (MR) characterization. As previously reported[14,17], the *b*-axis MR exhibits a distinct spin-valve effect near 6 T due to the transition from the AFM-b to a canted AFM (CAFM) state, resulting in a remarkable drop in MR.

Figure 2a displays the image of device S1. Electrodes 1 through 6 were patterned using optical photolithography. Owing to the approximately 1.5 μm thickness of the CRO crystal, the evaporated gold electrodes exhibit a discontinuity near the sample edges. This gap was bridged using platinum (Pt) deposited via FIB, as indicated by the light blue regions in Fig. 1a. The in-plane resistivity along the *b*-axis, $\rho_{bb}$, measured using contacts 2–3 with current along 1–4, is shown in Fig. 2b (blue curve). The measured $\rho_{bb}$ closely matches previously reported results[21,38,39], displaying a characteristic downturn at $T_N$, where the PM phase transitions to the AFM-a phase, followed by a sharp upturn at $T_S$, where the transition from AFM-a to AFM-b occurs. The sharp increase in $\rho_{bb}$ at $T_S$ is attributed to the opening of a pseudogap, as noted above. Unlike previous reports, however, $\rho_{bb}$ does not revert to a metallic temperature dependence below 30 K ($T^*$), likely due to a slight contribution from the out-of-plane resistivity component, $\rho_{cc}$, which arises from terraces on the sample surface and remains insulating below $T^*$.

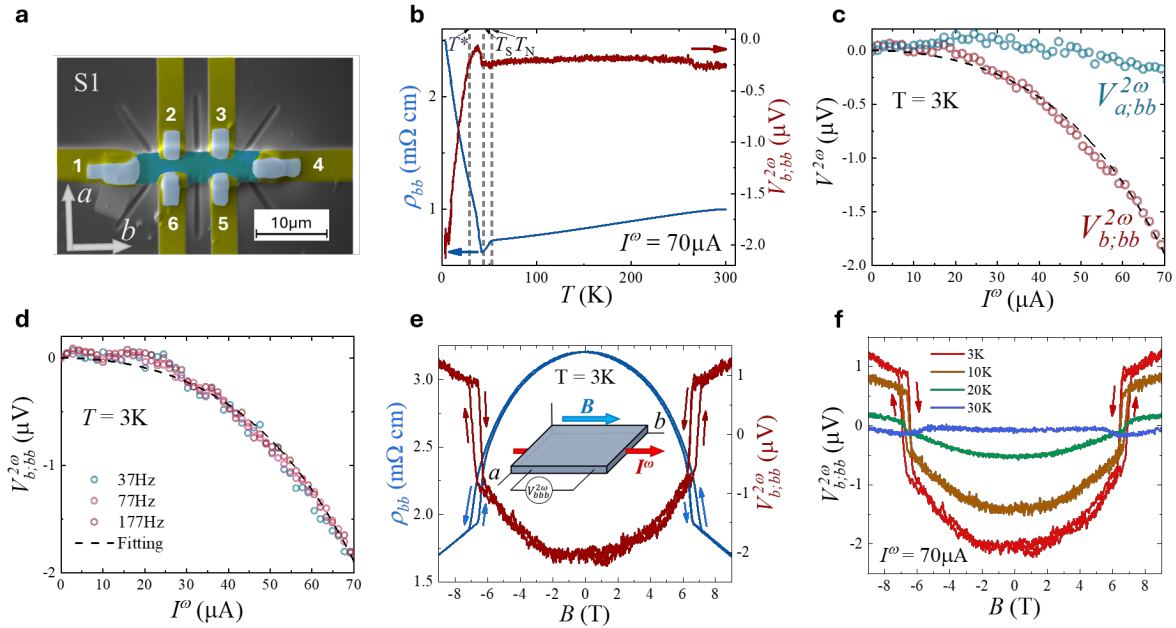

Figure 2. **Longitudinal nonlinear transport properties of CRO for the current applied along the $b$-axis**. **a** SEM image of the S1 device, with current applied along the $b$-axis. Gold electrodes (yellow) were patterned via thermal evaporation, and Pt contacts (light blue) were deposited by FIB to bridge gaps near the sample edges. **b** Temperature dependence of the in-plane resistivity along the $b$-axis and the second harmonic voltage $V_{b;bb}^{2\omega}$ under a 70µA current applied. Dashed lines indicate the characteristic temperatures $T^*$, $T_S$ and $T_N$ respectively. $V_{b;bb}^{2\omega}$ and $V_{a;bb}^{2\omega}$ denote second harmonic signals measured along the $b$- and $a$-axes, respectively, with current applied along the $b$-axis. **c** Second-harmonic voltages, $V_{b;bb}^{2\omega}$ and $V_{a;bb}^{2\omega}$, as functions of input ac current $I^\omega$ at 3K. **d** $V_{b;bb}^{2\omega}$- $I^\omega$ characteristics at different frequencies. Dashed lines in panels (c) and (d) represent quadratic+quartic fits to the data (Supplementary Note 3). **e** Magnetic field dependence of $\rho_{bb}$ and $V_{b;bb}^{2\omega}$ at 3K with a 70µA current. Inset to **e** is the schematic illustration of the sample orientation, with both the magnetic field and current applied along the $b$-axis. **f** Magnetic field dependence of $V_{b;bb}^{2\omega}$ at various temperatures. Arrows next to the curves in panels (e) and (f) indicate the direction of the field sweeps.

Through nonlinear transport measurements on device S1, we find compelling evidence that the AFM-b phase is associated with the $Pn2_1a$ structure. Figure 2c presents the longitudinal second-harmonic voltage response, $V_{b;bb}^{2\omega}$, measured between electrodes 2 and 3 under an AC current ($I^\omega$) applied along the $b$-axis at 3 K. $V_{b;bb}^{2\omega}$ is clearly nonzero, with its dependence on $I^\omega$ well described by a dominant quadratic term and a minor quartic correction, as shown by the dashed curve in Fig. 2c. The quartic term vanishes upon warming to 10 K (see supplementary Note 3). In addition, $V_{b;bb}^{2\omega}$ is independent of frequency (Fig. 2d). A similar longitudinal nonlinear response is also observed using electrodes 5 and 6 (see supplementary Note 4 and Fig. S3a & S3c). These features provide strong evidence for a finite NLR along the $b$-axis, indicating that AFM-b phase adopts the $Pn2_1a$ structure. This is further supported by the temperature evolution of $V_{b;bb}^{2\omega}$ across the magnetic phase transitions. As stated above, CRO exhibits two magnetic transitions upon warming, at $T_S$ and $T_N$, enabling us to track the nonlinear transport response across different magnetic phases. Figure 2b (red curve) presents $V_{b;bb}^{2\omega}$ as a function of temperature. Notably, as the temperature increases, the magnitude of $V_{b;bb}^{2\omega}$ gradually decreases, vanishing between $T^*$ and $T_S$, before rising to a small, nearly temperature-independent value for $T > T_S$. These results strongly suggest a correlation between the longitudinal nonlinear signal and magnetic ordering in CRO. The small but finite signal at high temperatures ($T > T_S$) is attributed to extrinsic effects such as the junction artifacts or thermoelectric contributions, as it is absent in other devices (see below).

As discussed earlier, the nonlinear response $V_{b;bb}^{2\omega}$ observed below $T_S$ should be attributed to the nontrivial band topology associated with a Weyl chain state that emerges in the AFM-b phase. This is further demonstrated from the coupling of $V_{b;bb}^{2\omega}$ with magnetism observed in magnetic field sweep measurements. As seen in the CRO's magnetic phase diagram in Fig. 1g, the magnetic field applied along the $b$-axis induces a transition from the AFM-$b$ to the CAFM state at approximately ~ 6 T for $T < 40$ K.

The CAFM state is characterized by a superposition of AFM and weak FM components, with spins canted 25° from the $b$-axis[14]. As noted above, this magnetic transition results in a spin-valve-like effect, i.e. the in-plane resistivity exhibits a steep drop due to significantly suppressed spin scattering. In our experiment, we observed a sharp decrease in $\rho_{bb}$ at ~ 7 T for the field applied along the long axis of device S1 (Fig. 2e), slightly higher than the previously reported spin-valve transition field ($H_{SV}$ ~ 6 T). This can be ascribed to a slight misalignment of the sample within the holder as $H_{SV}$ increases when the field is rotated away from the $b$-axis. From the data shown in Fig. 2e, it can be seen that $V_{b;bb}^{2\omega}$ shows a sharp change as well as a hysteresis near $H_{SV}$ in response to the AFM-b-to-CAFM transition, with its sign switching from negative to positive. These results indicate that longitudinal second-order current $j_\parallel^{2\omega}$ switches its sign as the AFM-b phase transitions to the CAFM phase. This information as well as its temperature dependence revealed in Fig. 2b & 2f has been schematically illustrated in the phase diagram in Fig. 1g. Such a coupling of $V_{b;bb}^{2\omega}$ with the magnetic transition can be well understood in terms of the variation of band topology across the magnetic transition revealed in our theoretical studies[25]. As mentioned earlier, in the AFM-$b$ state, CRO exhibits Weyl chains near the Fermi energy, which are protected by the underlying crystal and magnetic symmetries. However, upon transitioning to the CAFM state, spin canting changes the symmetry of the system, which has profound effects on the band structure, leading to the sign change of $j_\parallel^{2\omega}$.

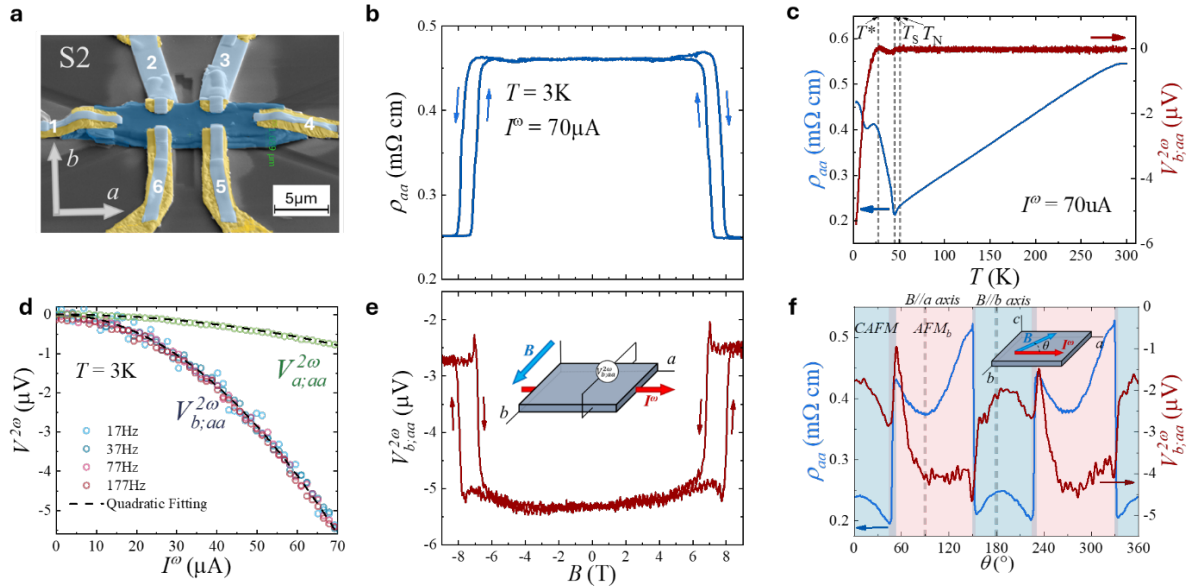

Figure 3. **Nonlinear Hall effect (NLHE) of CRO for the current applied along the a-axis**. **a** SEM image of a CRO (dark blue) device which allows current to be applied along the $a$-axis. **b, e** Magnetic field dependence of $\rho_{aa}$ (**b**) and $V_{b;aa}^{2\omega}$ (**e**) at 3 K, measured with a current of 70µA. The arrows next to the curves in **b** & **e** indicate the direction of magnetic field sweeps. Inset to **e**: Schematic of the sample orientation. **c** Temperature dependence of $\rho_{aa}$ and $V_{b;aa}^{2\omega}$ under a 70µA applied current. Dashed lines indicate the characteristic temperatures $T^*(=30K)$, $T_S(=48K)$ and $T_N(=56K)$. **d** The second-harmonic voltages, $V^{2\omega}_{b;aa}$ and $V^{2\omega}_{a;aa}$, as functions of input ac current $I^\omega$ at 3K, both showing a quadratic dependence on $I^\omega$. Open circles and dashed lines in **d** represent experiment data and quadratic fitting, respectively. **f** In-plane angular dependence of $\rho_{aa}$ and $V^{2\omega}_{b;aa}$ at 3K with a 70μA current. Inset to **f**: Schematic of the sample orientation, where current is applied along the *b*-axis and the field is rotated in-plane.

**Nonlinear transverse transport**

In addition to the NLR along the *b*-axis, the *Pn2₁a* structure phase is also predicted to exhibit a strong nonlinear Hall response, driven by the large QMD associated with Weyl chains near the Fermi energy (see Fig. 1f & Table 1), when the current is applied along the *a*- or *c*-axis. To probe the nonlinear Hall conductivity with $I^\omega//a$ ($\sigma_{b;aa}$), we fabricated a device (denoted S2) with its long axis along the *a*-axis. Figure 3a shows the image of this device, which was also fabricated on a single domain crystal. The temperature dependence of the in-plane resistivity along the *a*-axis ($\rho_{aa}$) and the second harmonic Hall voltage ($V^{2\omega}_{b;aa}$) measured on this device are presented in Fig. 3c. Like $\rho_{bb}$, $\rho_{aa}$ shows a downturn at $T_N$ and an upturn at $T_S$. However, near $T^*(=30K)$, $\rho_{aa}$ exhibits a peak, followed by an upturn, indicating that the contribution from $\rho_{cc}$, due to surface terraces, is smaller in S2 than in S1. At temperatures below 30 K, the magnitude of $V^{2\omega}_{b;aa}$ measured between voltage leads 3 and 5 increases sharply, reaching ~ 5 μV at 3 K, but vanishes above $T_S$ (see the red curve in Fig. 3c). This nonlinear Hall voltage shows a clear quadratic dependence on $I^\omega$ and is independent of frequency (Fig 3d), consistent with expectations for a second-order Hall response. This behavior is also reproduced in a second Hall channel measured between leads 2 and 6 (see Supplementary Note 4 and Fig. S3b & S3d). Meanwhile, the longitudinal second harmonic voltage ($V^{2\omega}_{a;aa}$) remains close to zero, consistent with DFT predictions in Table 1. The small residual values of $V^{2\omega}_{a;aa}$ are attributed to slight misalignment of the voltage leads.

Under an in-plane field along the *b*-axis, we observe sharp spin-valve transitions and hysteresis in both $\rho_{aa}$ (Fig.3b) and $V^{2\omega}_{b;aa}$ (Fig. 3e) near 7T, corresponding to the AFM-b to CAFM transition (Fig. 3b & 3e). Notably, $V^{2\omega}_{b;aa}$ decreases in magnitude in the CAFM state but does not change sign. This reduction suggests diminished topological contributions in the CAFM phase, as discussed above. In-plane field rotation measurements further support this trend (Fig. 3f), showing that the nonlinear signal is enhanced in the AFM-b state and suppressed in the CAFM state. During the field rotation driven magnetic transitions, we also observe sharp peaks in $V^{2\omega}_{b;aa}$ (Fig. 3f) at the phase's boundaries, consistent with signatures of metamagnetic transition. These features possibly arise from spin scattering at spin domain boundaries due to noncollinear canting of magnetic moments[40]. The NLHE driven by an *a*-axis current is also reproduced in another microscale device (S3) (see Supplementary Fig. S4), as well as in a millimeter-scale device (S4) containing multiple domains (see Supplementary Fig. S5), demonstrating the robustness of this nonlinear Hall response across different device geometries and domain configurations.

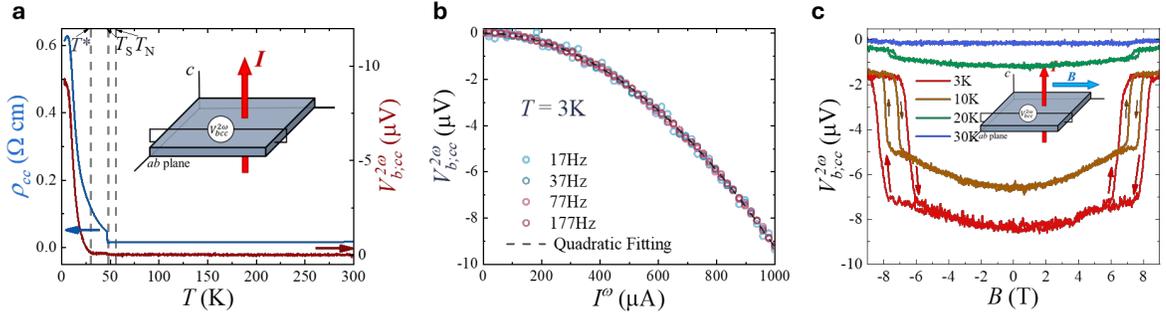

Figure 4. **NLHE of CRO for the current applied along the c-axis. a** Temperature dependence of the *c*-axis linear resistivity $\rho_{cc}$ and the second harmonic in-plane Hall voltage $V_{b;cc}^{2\omega}$ with 1mA current applied along the *c*-axis. Dashed lines indicate the characteristic temperatures of $T^*$, $T_S$ and $T_N$. **b** $V_{b;cc}^{2\omega}$ as a function of input current $I^\omega$ at 3K. **c** Magnetic field dependence of $V_{b;cc}^{2\omega}$ under a 1mA current applied along the c-axis at various temperatures. Arrows in **c** indicate the directions of field sweeps.

To confirm the NLHE induced by the *c*-axis current (i.e. $\sigma_{b;cc}$ in Table 1), we also performed measurements on single crystals with the current applied along the *c*-axis. In this experiment, we could not make a device on a single domain crystal, instead, we used a CRO crystal with the lateral dimensions of ~1 mm and the electrodes being attached using silver epoxy (denoted S5). Although such a crystal has domains, we expect to observe in-plane nonlinear Hall response along the sample edge (i.e. the *a*/*b*-axis), as shown in the insert to Fig. 4a. From the measurements on device S5, we indeed observed in-plane second harmonic Hall response $V_{b;cc}^{2\omega}$ (see the red curve in Fig. 4a where we also present the *c*-axis resistivity $\rho_{cc}$ data measured on another sample). Like $V_{b;aa}^{2\omega}$, the magnitude of $V_{b;cc}^{2\omega}$ sharply increases below $T^*$, and vanishes above $T_s$. $V_{b;cc}^{2\omega}$ also exhibits a clear quadratic dependence on $I^\omega$ and remains frequency independent (Fig. 4b). In-plane magnetic field scans show similar sharp transitions and hysteresis in $V_{b;cc}^{2\omega}$ near the spin-valve transition field (Fig. 4c), with a reduction in magnitude in the CAFM state. For comparison, we also conducted transverse nonlinear measurements on device S1 (Fig. 2a). Since the current in this device is applied along the *b*-axis, NLHE is not theoretically expected for this measurement configuration (see Table 1). In experiments, the measured transverse second harmonic response $V_{a;bb}^{2\omega}$ is indeed negligibly small (Fig. 2c) and its small value is caused by the voltage leads' misalignment, consistent with the theoretical prediction of the absence of NLHE for this measurement configuration.

**Discussion and outlook**

One may ask whether the NLR and NLHE observed in the AFM-*b* phase of CRO could arise from extrinsic contributions such as junction artifacts, thermoelectric effects, or disorder-induced scattering. This possibility can be excluded based on the consistent behavior observed across five independent devices studied in this work. In four of these devices, both the NLR and NLHE emerge only below the magnetic transition temperature $T_S$ and are strongly enhanced below $T^*$, but vanish above $T_S$ (Figs. 2b, 3c, and 4a), indicating their intrinsic link to the magnetic phase transition. Furthermore, both signals are suppressed across the field-induced AFM-*b* to CAFM transition, showing a parallel response to the change of magnetic ordering (Figs. 2e and 3e). Such robust coupling to magnetic transitions is not expected for nonlinear signals of extrinsic origin (*e.g.*, disorder scattering [11,41,42]), which is insensitive to the spin order or magnetic transition. There is only one exception: device S1 shows a weak, constant nonlinear response above $T_S$ (Fig.2b), likely due to a trivial source, as this feature is absent in the other four devices - S2 (Fig. 3), S3 (Fig. 4), and S4 and S5 (Supplementary Figs. S4–5). Therefore, the observed longitudinal NLR along the *b*-axis (Fig. 2), as well as the NLHE for $I^\omega$//*a* (Fig. 3) and $I^\omega$//*c* (Fig. 4) in the AFM-*b* phase, are intrinsic in nature and originate from nontrivial band topology, as illustrated in Fig. 1f. The agreement of these observations with the nonlinear conductivity tensor components predicted for the *Pn2₁a* structure (Table 1) supports the conclusion that the magnetic transition at $T_S$ - where the magnetic moments of Ru rotate from the *a*-axis ($T_S < T < T_N$) to the *b*-axis ($T < T_S$) - induces subtle structural distortions undetectable by neutron scattering. These distortions break $\tau T$ symmetry, as shown in Fig. 1a-b, and are essential for the emergence of the observed nonlinear responses.

Our experimental data also show that the nonlinear Hall response is significantly stronger than the longitudinal nonlinear response along the *b*-axis. Under an applied current of 70 μA, the second-harmonic Hall voltage measured in device S2 reaches approximately 5.5 μV at 3 K (Fig. 3d), more than twice the longitudinal second-harmonic voltage observed in device S1 (Fig. 2d). Since the crystals used in S1 and S2 have similar dimensions, this difference is intrinsic and aligns with theoretical prediction [ref. 25]. To enable a quantitative comparison with theory, we estimate the nonlinear conductivities using the relation $\sigma_{i;jj}^{2\omega} = (V_{i;jj}^{2\omega} l^3)/(I_j^2 R_{jj}^3 w^2 d)$ [ref.31] , where *l*, *w*, and *d* are the sample length, width and thickness, respectively; $I_j$ is the applied current and $R_{jj}$ is the linear resistance. Using measured values of $\rho_{aa} = 460$ μΩ.cm and $\rho_{bb} = 2.5$ mΩ.cm at 3K, we obtain $\sigma_{b;bb}^{2\omega} = 0.25 \frac{A}{V^2}$ for device S1 and $\sigma_{b;aa}^{2\omega} = 150 \frac{A}{V^2}$ for device S2. Theoretical calculation[25] yield $\sigma_{b;bb}^{2\omega} \sim 0.05 \frac{A}{V^2}$ and $\sigma_{b;aa}^{2\omega} \sim 10 \frac{A}{V^2}$ near the charge neutral point. Although the experimental values are approximately one-order of magnitude larger, the ratio between $\sigma_{b;bb}^{2\omega}$ and $\sigma_{b;aa}^{2\omega}$ is consistent between theory and experiment. This trend is reminiscent of similar behavior observed in the magnetic topological insulator MnBi$_2$Te$_4$[31]. The larger experimental values of $\sigma_{b;bb}^{2\omega}$ and $\sigma_{b;aa}^{2\omega}$ may be associated with contact misalignment and surface terraces on the crystals. Similarly, using the measured value of $\rho_{cc} = 0.62\,\Omega\cdot\text{cm}$, $\sigma^{2\omega}_{b;cc}$ is estimated to be $0.0029\,\frac{A}{V^2}$ for device S5, one order of magnitude smaller than the theoretical value of $\sigma^{2\omega}_{b;cc} \sim 0.05\,\frac{A}{V^2}$. We attribute this lower experimental value to multidomain structure, since both twin domains and 180° domains contribute oppositely to the net nonlinear signal. Additionally, the vanishing $\sigma^{2\omega}_{a;bb}$ (Fig. 2c) and $\sigma^{2\omega}_{a;aa}$ (Fig. 3d) components further support the agreement between our measured nonlinear conductivity tensor and theoretical expectations. Most importantly, the observation of NLR ($\sigma^{2\omega}_{b;bb}$) in CRO provides direct evidence of the concurrent inversion and $\tau T$ symmetry breaking - conditions uniquely satisfied in the low temperature AFM-*b* phase that adopts *Pn2₁a* space group (Fig. 5). This assignment not only resolves the missing symmetry-breaking in the phase transition but also establish the altermagnetic ground state of CRO for the first time.

Based on our theoretical studies[25], both the longitudinal nonlinear response and the NLHE of the AFM-*b* phase should be dominated by QMD as indicated above. In theory, the sign of QMD is dependent on the Néel vector direction and can be reversed if the Néel vector can be reversed by controlling the direction of magnetic field sweep. The sign switching of QMD would lead to the sign reversal of the nonlinear signal. This was previously observed in few layers magnetic topological insulator MnBi$_2$Te$_4$ [33,34,43–45], where the field driven Néel vector switching is enabled by reduced interlayer coupling, low anisotropy energy, broken inversion symmetry and a strong coupling between magnetic and electronic degrees of freedom. In our field sweep measurements on CRO, we find the nonlinear signal remain the same sign upon positive and negative field sweep (see Fig. 2e, 3e, 4c). This is because that in bulk CRO both the exchange energy and magnetic anisotropy energy are large such that magnetic field alone cannot reverse the Néel vector.

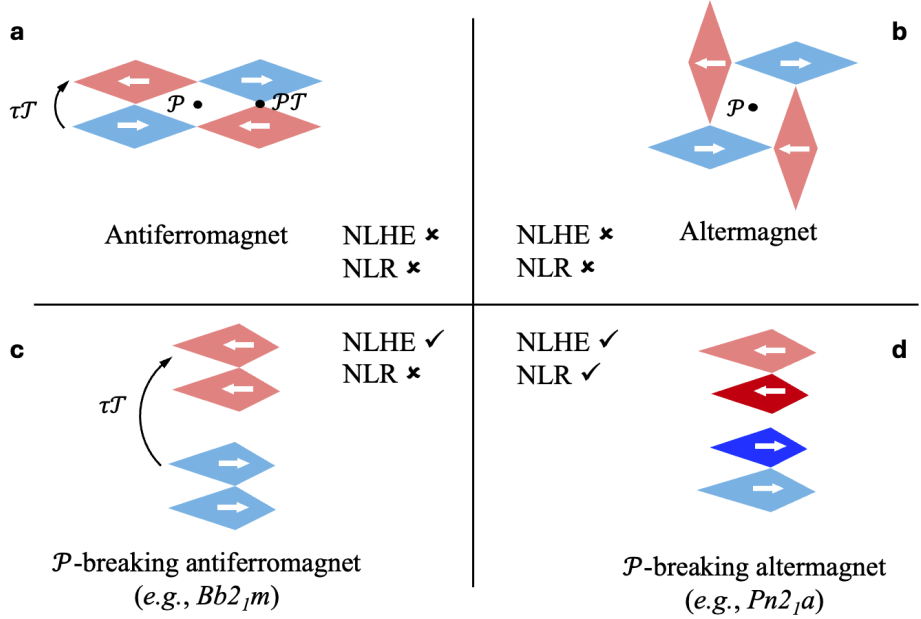

Figure 5. **Schematics of antiferromagnets and altermagnets with different nonlinear responses**. Opposite spin-sublattices are related by $\tau T$ or $PT$-symmetry in antiferromagnets (**a,c**) or rotation/mirror symmetry in altermagnets (**b,d**). In $P$-symmetric systems (**a,b**), both nonlinear Hall effect (NLHE) and nonlinear resistance (NLR) vanishes. In $P$-breaking case, NLHE may exist antiferromagnets or altermagnets (**c,d**) while NLR only occurs in altermagnets (*e.g.*, the $Pn2_1a$ phase of $Ca_3Ru_2O_7$).

As shown in Figs. 2b, 3c, and 4a, both the longitudinal nonlinear response along the *b*-axis and the NLHE are significantly enhanced below $T^* = 30$ K. This enhancement can be attributed to a pronounced reconstruction of the Fermi surface at low temperatures. Previous ARPES studies on CRO[20,22,46] have revealed that, below 30 K, the electronic structure becomes highly anisotropic, with bands crossing the Fermi level only near the $M_a$ and $M_b$ points, and the emergence of a Dirac cone around $M_a$ [22]. Earlier first-principles calculations further predicted the presence of Weyl nodes embedded within this Dirac cone at $M_a$ [26]. Our recent calculations [25] indicate that these Weyl points extend and form Weyl chains confined within glide mirror planes. However, resolving these fine topological features via ARPES remains challenging due to the narrow energy window of the Weyl bands (~10 meV), and the need for a clean cleavage along the *ac*-plane to visualize the associated surface states. Although the $Pn2_1a$ structure leads to presence of an altermagnet, the associated nonrelativistic spin splitting is too small to be resolved by ARPES. Moreover, the anomalous Hall effect is symmetry-forbidden due to the presence of two glide mirrors. Consequently, unlike in other known altermagnets[47–52], the altermagnetic state in CRO cannot be detected using conventional probes such as ARPES or the anomalous Hall effect. Demonstration of the $\tau T$ symmetry breaking via nonlinear transport measurements is the only approach to prove the altermagnetic state of CRO as discussed above (see Figure 5). This approach can possibly be employed to identify a broad class of altermagnetic materials with broken inversion symmetry, as to be discussed below.

The common AFM materials have $\tau T$ and/or $PT$ symmetry, where $P$ stands for the spatial inversion (Fig. 5a & 5c), while altermagnets are a unique group of AFMs that lacks both $\tau T$ and $PT$ (Fig. 5b & 5d). Most known altermagnets are inversion symmetric and thus forbid the second-order nonlinear transport (Fig. 5b) while inversion-breaking altermagnets can exhibit nonlinear effects (Fig. 5d). Therefore, the nonlinear transport can serve as a sensitive probe to differentiate the altermagnet from an ordinary AFM with $\tau T$ and distinguish two types of altermagnets (inversion-breaking or inversion symmetry), as shown in Fig. 5b & 5d. Beyond the specifics of CRO, our findings underscore the broader significance of nonlinear transport as a universal and sensitive probe of subtle symmetry breaking. Across a wide class of systems, minute symmetry differences—whether driven by magnetic transitions, non-stoichiometric compositions, or mixed-valence states—often give rise to profound changes in electronic topology. However, distinguishing these nearly degenerate structural or magnetic configurations remains a central challenge, particularly when conventional probes such as diffraction or ARPES are limited by spatial resolution. In this context, the nonlinear response, which directly couples to symmetry-breaking components of the quantum geometric tensor, provides a robust bulk-sensitive signature. Thus, our work highlights nonlinear transport as a definitive diagnostic across a range of quantum materials where conventional methods fall short, enabling direct access to hidden symmetry breaking and associated topological phenomena.

In summary, we have demonstrated intrinsic quantum metric-induced nonlinear transport in the strongly correlated oxide $Ca_3Ru_2O_7$, marking the first observation of such phenomena in oxide materials. Our nonlinear transport measurements conclusively resolve the longstanding ambiguity regarding CRO's ground state structure, supporting the lower symmetry $Pn2_1a$ phase, and establishing the altermagnetic state. The remarkable sensitivity of the nonlinear transport signal to the subtle symmetry change caused by the spin reorientation transition underscores its effectiveness as a powerful diagnostic tool in exploring hidden symmetries and topological states in quantum materials. Future studies utilizing nonlinear transport promise deeper insights into hidden symmetry breaking and topology in quantum materials.

**Methods:**

Single crystals of $Ca_3Ru_2O_7$ (CRO) were synthesized using the floating zone method. Microscale devices were fabricated via photolithography and focused ion beam (FIB) technique using an FEI Scios 2 dual-beam scanning electron microscope (SEM). Gold electrodes were patterned using standard photolithography techniques. Platinum electrical contacts were deposited in situ within the SEM chamber. A cleaning step with ion beam was performed to remove spurious debris and eliminate potential electrical shorts. The FIB-fabricated devices were placed on $SiO_2$ substrates to avoid spurious nonlinear signals, as previously indicated when Pt is deposited on Si substrates[53].

Nonlinear transport measurements were conducted using standard lock-in detection methods. Alternating current (AC) and direct current (DC) were supplied by a Keithley 6221 and a Keithley 6220 precision current source, respectively. Corresponding voltage signals were measured using a Stanford Research Systems SR860 lock-in amplifier for AC and a Keithley 2182A nanovoltmeter for DC measurements.

Out-of-plane contact fabrication on FIB-patterned devices require additional manipulation steps that can introduce surface damage and affect measurement reliability, hence out-of-plane measurements were performed on millimeter-scale devices. A small sample dimension is essential for obtaining a measurable second harmonic signal; therefore, we made two devices: a smaller device S3 with four contacts for second harmonic measurements and a larger device, S4, with six contacts for linear transport measurements. The principle in-plane axes of S3 were determined using spin valve measurements; although, the presence of multiple twin domains cannot be ruled out.